\documentclass{svjour2}                    
\smartqed  
\usepackage{graphicx}
\usepackage{dcolumn}
\usepackage{bm}
\usepackage{graphicx}
\usepackage{color} 
\usepackage{float}
\DeclareGraphicsExtensions{. jpg,. pdf, . mps, . png, . eps, . ps, . EPS}
\begin{document}

\def\be{\begin{equation}}
\def\ee{\end{equation}}

\def\bc{\begin{center}}
\def\ec{\end{center}}
\def\bea{\begin{eqnarray}}
\def\eea{\end{eqnarray}}
\newcommand{\avg}[1]{\langle{#1}\rangle}
\newcommand{\Avg}[1]{\left\langle{#1}\right\rangle}
\newcommand{\cor}[1]{\textcolor{red}{#1}}

\title{Percolation  on  interdependent networks with a fraction  of  antagonistic interactions
}


\author{Kun Zhao         \and
        Ginestra Bianconi 
}


\institute{Kun Zhao \at
              Department of Physics, Northeastern University, Boston,
Massachusetts 02115 USA \\
           \and
           Ginestra Bianconi \at
                            School of Mathematical Sciences, Queen Mary University of London, London E1 4NS, United Kingdom \\
                            \email{ginestra.bianconi@gmail.com}
}

\date{Received: date / Accepted: date}

\maketitle

\begin{abstract}
 Recently, the percolation transition has been characterized on interacting networks both  in presence of interdependent and antagonistic interactions.
 Here we characterize the phase diagram of the percolation transition in two Poisson interdependent networks with a percentage $q$ of antagonistic nodes. We  show that this system can present  a bistability of the steady state  solutions, and both  first, and second order phase transitions. In particular, we observe a bistability of the solutions in some regions of the phase space also for a small fraction of antagonistic interactions $0<q<0.4$.
Moreover, we   show  that a fraction $q>q_c=2/3$ of antagonistic interactions is necessary to strongly reduce  the region in  phase-space in which both networks are percolating. 
This last result suggests that interdependent networks are robust to the presence of antagonistic interactions.
Our approach can be  extended to multiple networks, and to complex boolean rules for regulating the percolation phase transition.
\keywords{Percolation \and  Antagonistic interactions \and Interdependent networks}
\end{abstract}

\section{Introduction}

Percolation \cite{Havlin1,MolloyR,Havlin2} is one of the most relevant critical phenomena \cite{crit,Dyn,isi_g,Doro_isi,Zecchina_isi,Bradde_PRL,VespignaniES,MunozES,TorozckaiES,Motter_h,Motter_sw,Synchr,revSyn,TJamming,Moreno,Cong,QTIM,BH,JSTAT}
that can be defined on a complex network.
Investigating the properties of percolation on single network reveals the essential role played by  the  topology of the network  in determining the network   robustness \cite{Havlin1,MolloyR}.
Recently, large attention has been paid  to the study of the percolation transition on complex networks and surprising new phenomena have been observed. On one side, new results have shown that the percolation can be retarded and sharpened by the Achlioptas process \cite{Raissa,Fortunato_exp,Dorogovtsev_exp,Riordan}. On the other side, it has been shown that when considering interacting networks, the percolation transition can be first order \cite{Havlin_intPRL,Havlin_intPNAS,Grassberger}.
This last result is extremely interesting because a large variety of networks are not isolated but are strongly interacting \cite{Havlin_intPRL,Havlin_intPNAS,Grassberger,Havlin_int1,Vespignani,Dorogovtsev,Yamir1,Yamir2,Jesus,Ivanov}.  In these systems one network function depends on the operational level of the other networks. Examples of investigated interacting networks go from infrastructure networks as the power-grid \cite{Havlin_int1} and the Internet to interacting biological networks in physiology \cite{Ivanov}. 
Nodes in interacting networks can be interdependent, and in this case  the function or activity of a node depends on the function of the activity of the linked nodes in the others networks. 
Recent results have shown that interdependent networks are more fragile than single networks  \cite{Havlin_int1,Vespignani} with serious  implications   that these results have on an increasingly interconnected world.

Nevertheless, in interacting networks we might  also  observe antagonistic interactions. If two nodes have an antagonistic interaction, the functionality, or activity, of a node in a network is incompatible with the functionality, of the other node in the interacting network. This new possibility \cite{PerAnt}, opens the way to introduce in the interaction networks antagonistic interactions that generate a bistability of the solutions.

In this paper we introduce  a fraction of antagonistic interactions in two  otherwise interdependent networks and we study the interplay between interdependencies and antagonistic interactions.
We consider this problem in a simplified settings by   looking  at two interacting Poisson networks.
 For two Poisson networks with exclusively interdependent  interactions, the steady state of the percolation dynamics has a large region of the phase diagram in which both networks are percolating.
In these system, a fraction $q>q_c=2/3$ of antagonistic interactions is necessary in order to significantly reduce the region in phase-space in which both networks are percolating.
This show that interdependent networks display a significant  robustness in presence of antagonistic interactions, and that also a minority of interdependent nodes is enough to sustain  two percolating networks.

The paper is structured as follow: in section II we will review the theory of percolation in single random networks, in section III we will review the theory of percolation in interdependent networks, in section IV we will characterize  the percolation phase diagram of two Poisson networks with purely antagonistic interactions, in section V we will characterize the percolation phase diagram in networks with a fraction $q$ of antagonistic nodes and a fraction $1-q$ of interdependent nodes, finally in section VI we will give the conclusions.

\section{Percolation on single network}

Over the past ten years great attention has been paid to the percolation transition on single networks.
The percolating cluster in a single Poisson network emerges at a second order phase transition when the average degree of the network is $\avg{k}=1$. Nevertheless, this result can change significantly for networks with different degree distributions.

In order to solve the percolation problem in a random network with degree distribution $p_k$ we make use of the generating functions $G_0(x),G_1(x)$ defined as in the following:
\begin{eqnarray}
G_1(x)&=&\sum_k \frac{k p_k}{\avg{k}}x^{k-1}\nonumber \\
G_0(x)&=&\sum_k p_k x^k,
\label{Gen}
\end{eqnarray}
We indicate by $S$ the probability that a node is part of the percolating cluster, and by $S^{\prime}$ the probability that following a link we reach a node that belongs to the percolating cluster.
Each node of the network belongs to the percolating cluster of the network  if at least one of its  links brings to a node which is part of the percolating cluster of the network.
Expressing this observation in terms of $S$ and $S^{\prime}$, we obtain the relation 
\begin{eqnarray}
S=[1-G_0(1-S^{\prime})].
\label{single_g0}
\end{eqnarray}
Moreover,  the probability $S^{\prime}$ can be found by solving the following recursive equation valid on a locally tree-like network,
\begin{eqnarray}
S^{\prime}=[1-G_1(1-S^{\prime})].
\label{single_rec}
\end{eqnarray}
These equations are the well known equations for the percolation transition on single network \cite{Havlin1,MolloyR} with given  degree distribution.
Equation $(\ref{single_rec})$ has a non trivial solution $S^{\prime}>0$ which emerges continuously at  a second order phase transition when
\begin{equation}
\left.\frac{d G_1(x)}{dx}\right|_{x=1}=\frac{\avg{k(k-1)}}{\avg{k}}=1.
\label{per_cond0}
\end{equation}
The percolating cluster will be present in the network as long as
\begin{equation}
\frac{\avg{k(k-1)}}{\avg{k}}> 1.
\label{per_cond}
\end{equation}
Therefore for Poisson networks we have derived  that the percolation condition Eq. $(\ref{per_cond})$ prescribes that   the average connectivity of the network $z=\avg{k}=\avg{k(k-1)}/\avg{k}$ should be  greater than one, i.e.  we must have $z> 1$ for the network to be percolating.
For scale-free networks, with power-law degree distribution $p(k)\propto k^{-\gamma},$  the percolation condition Eq.  $(\ref{per_cond})$ implies that the network, as long as the power-law exponent $\gamma\leq3$, is always percolating  in the thermodynamic limit $N\to \infty$. Indeed in this case  the   second moment of the degree distribution is diverging with the network size, i.e. $\avg{k^2}\to \infty$ for  $N \to \infty$ . This is a crucial result in complex networks theory and implies that scale-free networks with exponent $\gamma\leq 3$ are more robust than any other network with finite second moment of the degree distribution, i.e. with  $\avg{k^2}<\infty$.

\section{Percolation on two interdependent networks}

In this section we will review the theory of percolation on two interdependent networks following the approach developed by Son et al. \cite{Grassberger}.
We will assume that the two networks are called network A and network B and that both networks have the same number of nodes $N$. In other words our interacting networks constitute a multiplex.
In fact each node is represented in both networks. A node $i$ belongs to the percolating cluster of the interdependent networks if the two following condition are met 
\begin{itemize}
\item
{\it (i)} at least  one of the neighbour nodes of $i$ in network A belongs to  the percolating cluster of the interdependent networks.\\

\item
{\it (ii)} at least  one of the neighbour nodes of $i$ in network B belongs to  the percolating cluster  of the interdependent networks.
\end{itemize}
If we denote by $S$ the probability that a node belongs to the percolating cluster of two interdependent networks and by $S^{\prime}$ the probability that following a link we reach a node in the percolating cluster of the interdependent networks we have 
\begin{eqnarray}
S&=&[1-G_0^A(1-S^{\prime}_A)][1-G_0^B(1-S^{\prime})_B] 
\label{int_g0}
\end{eqnarray}
The recursive equations for  $S^{\prime}_A$ and $S^{\prime}_B$, valid on a tree-like random network  are given by
\begin{eqnarray}
S^{\prime}_A&=&[1-G_1^A(1-S^{\prime}_A)][1-G_0^B(1-S^{\prime}_B)]\nonumber \\
S^{\prime}_B&=&[1-G_1^B(1-S^{\prime}_B)][1-G_0^A(1-S^{\prime}_A)].
\label{int_rec}
\end{eqnarray}
The interesting new result is that now the percolation transition can be also first order \cite{Havlin_int1,Havlin_intPRL,Havlin_intPNAS,Grassberger} as the following paragraphs show in simple cases.

\subsection{Two Poisson networks with equal average degree $z$.}
The percolation on interdependent  networks was first studied in \cite{Havlin_int1,Havlin_intPRL} and then further characterized in \cite{Grassberger}.
A relevant example  of interdependent networks  is represented by two Poisson networks with the same average degree $z=\avg{k}_A=\avg{k}_B$.
In the case of  Poisson networks the generating functions are given by $G_0^A(x)=G_1^A(x)=G_0^B(x)=G_1^B(x)=e^{z(x-1)}$. Therefore we have a relevant simplification of our Eqs. $(\ref{int_g0})-(\ref{int_rec})$ because  $S=S^{\prime}_A=S^{\prime}_B$.
The equation for $S$ (Eqs. $(\ref{int_rec}),(\ref{int_g0})$) now reads
\begin{equation}
S=\left[1-e^{-zS}\right]^2.
\label{int_simple}
\end{equation}
By defining $g(S)=S-[1-e^{-zS}]^2$ the Eq. $(\ref{int_simple})$ is equivalent to $g(S)=0$.
This equation has always the solution $S=0$ but as a function for  $z=z_c$ the curve $g(S)$ is tangential to the $x$ axis and another non trivial solution emerges.

\begin{figure}
\begin{center}
\includegraphics[width=.6\columnwidth]{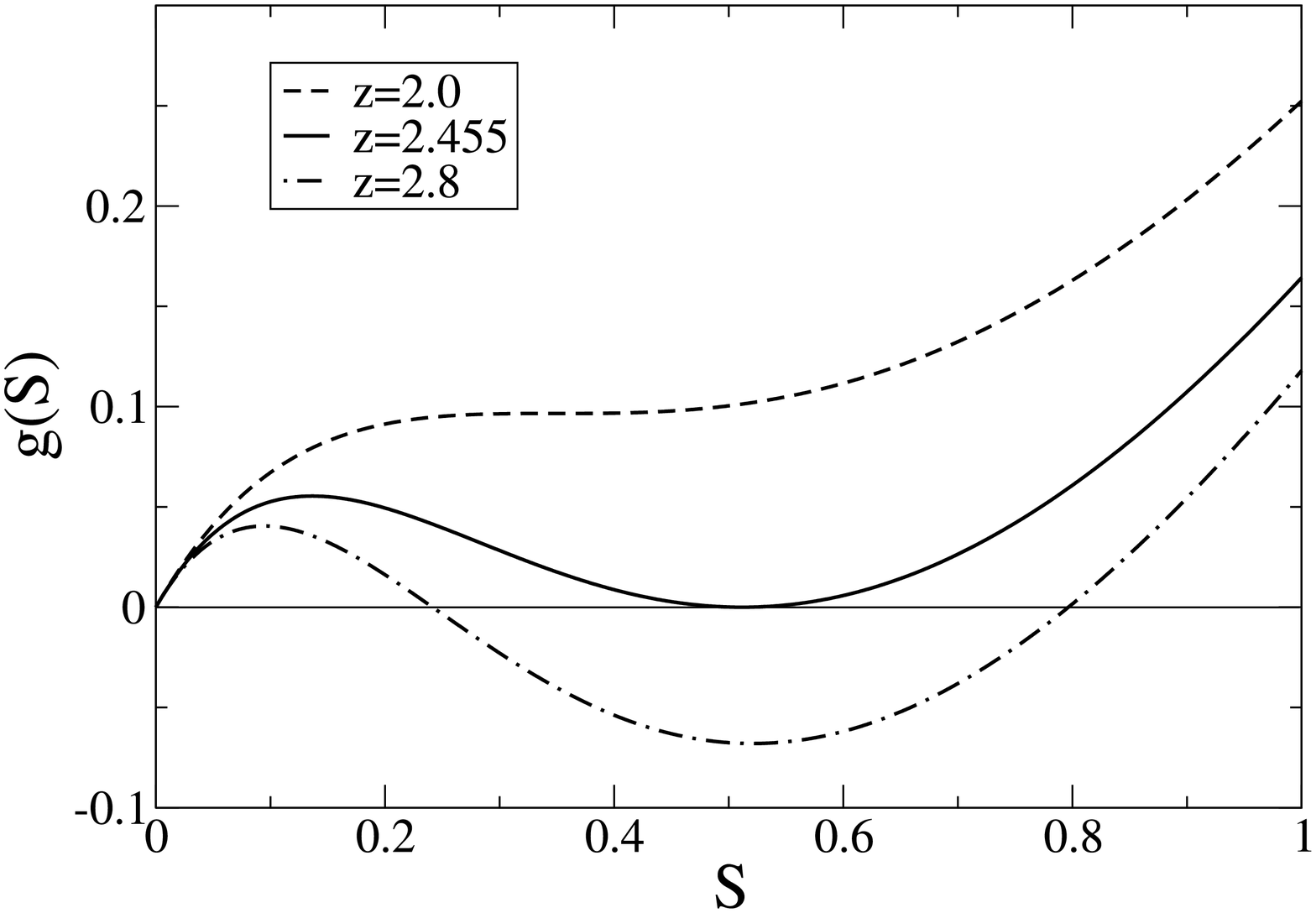}
\end{center}
\caption{Plot of the function $g(S)$ for different values of average connectivity $z$. At $z=z_c=2.455\ldots$ a new non-trivial solution of the function $g(S)=0$ indicates the onset of a first order phase transition. }
\label{first_order}
\end{figure}

The point  $z=z_c$ can be found by imposing the condition 
\begin{eqnarray}
g(S)&=&0,\nonumber \\
\frac{dg(S)}{dS}&=&0,
\end{eqnarray}
identifying the point when the function $g(S)$ is tangential to the $x$ axis.
Solving this system of equations we  get   $z=z_c=2.455407\dots$ and $S_c=0.511699\dots$.
In Figure $\ref{first_order}$ we show a plot of the function $g(S)$ for different values of the average connectivity of the network $z$ below and above the first order phase transition $z=z_c$.
For $z<z_c$ the only solution to Eq. $(\ref{int_simple})$ is $S=0$, for $z= z_c$ a new non trivial solution emerge with $S= S_c$. Therefore at $z=z_c$ we observe a phase transition of the first order in the percolation problem.

\begin{figure}
\begin{center}
\includegraphics[width=.6\columnwidth]{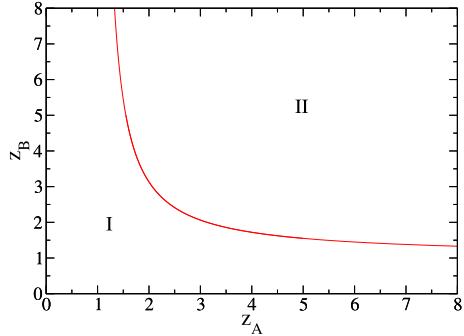}
\end{center}
\caption{Phase diagram of two interdependent Poisson networks with average degree $z_A$ and $z_B$ respectively. In region I we have $S=0$, in region II we have $S>0$ and the critical line indicates the points where the first-order transition occurs.}
\label{inter_ER_ER}
\end{figure}

\subsection{Two Poisson networks with different average degree $z_A\neq z_B$}

Another important example of interdependent networks is the the case investigated in \cite{Grassberger} of two Poisson networks with different average degrees $\avg{k}_A=z_A$ and $\avg{k}_B=z_B$.
In the case of  Poisson networks the generating functions are given by $G_0(x)=G_1(x)=e^{z(x-1)}$. Therefore we have a relevant simplification of our Eqs. $(\ref{int_g0})-(\ref{int_rec})$ because  $S=S^{\prime}_A=S^{\prime}_B$.
The equation for $S$ (Eqs. $(\ref{int_rec}),(\ref{int_g0}))$ now reads

\begin{equation}
\Psi(S)=S-(1-e^{z_AS})(1-e^{z_BS})=0
\end{equation}
The discontinuous phase transition can be found by imposing the following conditions 
\begin{eqnarray}
\Psi(S)&=&0,\nonumber \\
\frac{d\Psi(S)}{dS}&=&0.
\end{eqnarray}
In Figure $\ref{inter_ER_ER}$ we plot the phase diagram of the percolation process on these two interdependent networks.
In this phase diagram we have a large region (Region II) in which both networks are percolating ($S>0$) and we observe a first order percolation phase transition on the critical line of the phase diagram.
\section{ Percolation on two antagonistic networks}
In a recent paper, we have  introduced antagonistic interactions in the percolation of two interacting networks \cite{PerAnt}.
As in the case of interdependent networks we consider two networks of $N$ nodes. We call the networks, network A and network B respectively and every node $i$ is represented in both networks, i.e. the networks form a multiplex.
The difference with the case of interdependent network is that  if a node $i$ belongs to the percolating cluster of  on one network it cannot belong to the percolating cluster of   the other one.
A node $i$ belongs to the percolating cluster  of network A (network B) if the following two conditions are met:
\begin{itemize}
\item {\it (i)} at least one node reached by following the links incident to node $i$ in network A (network B) belongs to the percolating cluster in network A (network B);
\item {\it (ii)} none of the nodes reached by following the links incident to node $i$ in network B (network A) belongs to the percolating cluster in network B (network A).
\end{itemize}
If we indicate by $S_{A} (S_B)$ the probability that a node in network A (network B) belongs to the percolating cluster in network A (network B), and if we indicate by $S_A^{\prime}(S_B^{\prime})$ the probability that following a link in network A (network B) we reach a node in the percolating cluster of network A (network B), we have 
\begin{eqnarray}
S_A&=&[1-G_0^A(1-S_A^{\prime})]G_0^B(1-S_B^{\prime})\nonumber\\
S_B&=&[1-G_0^B(1-S_B^{\prime})]G_0^A(1-S_A^{\prime})
\label{g0int}
\end{eqnarray}
\begin{figure}
\center
\includegraphics[width=0.8\columnwidth]{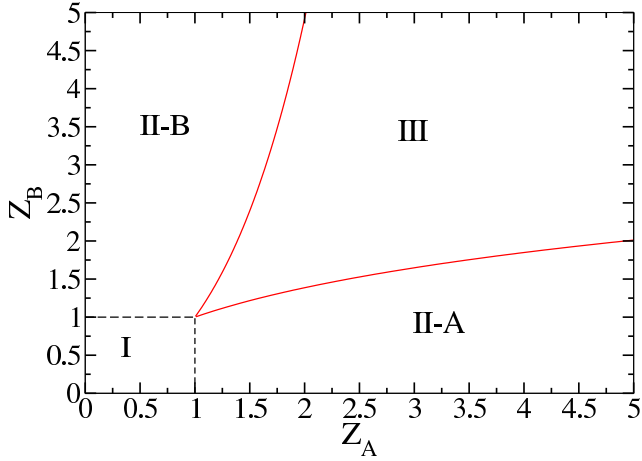}
\caption{Phase diagram of two antagonistic Poisson networks with average degree $z_A$ and $z_B$ respectively. In region I the only stable solution is the trivial solution $S_A=S_B=0$. In region II-A we have only one stable solution $S_A>0, S_B=0$, Symmetrically in region II-B we have only one stable solution $S_A=0,S_B>0$. On the contrary in region III  we have two stable solutions $S_A>0,S_B=0$ and $S_A=0,S_B>0$ and we observe a bistability of the percolation steady state solution.}
\label{ant_ER_ER}
\end{figure}
At the same time, in a random graph with local tree structure  the probabilities   $S_A^{\prime}$ and $S_B^{\prime}$  satisfy the following recursive equations
\begin{eqnarray}
S_A^{\prime}&=&[1-G_1^A(1-S^{\prime}_A)]G_0^B(1-S_B^{\prime})\nonumber\\
S_B^{\prime}&=&[1-G_1^B(1-S^{\prime}_B)]G_0^A(1-S_A^{\prime}).
\label{rec}
\end{eqnarray}

\subsection{Two Poisson networks}
We consider the case of two Poisson networks with  average connectivity $\avg{k}_A=z_A$ and $\avg{k}_B=z_B$.

In this case,  the generating functions take the simple expression $G_1^A(x)=G_0^A(x)=e^{-z_A(1-x)}$ and   $G_1^B(x)=G_0^B(x)=e^{-z_B(1-x)}$.
Therefore, taking into consideration Eqs.$(\ref{g0int})$ and Eqs. $(\ref{rec})$ we have $S_{A}^{\prime}=S_{A}$ and $S_{B}^{\prime}=S_{B}$.
Moreover   Eqs.(\ref{rec}) take the following form:
 \begin{eqnarray}
S_A&=&(1-e^{-z_AS_A})e^{-z_BS_B}\nonumber\\
S_B&=&(1-e^{-z_BS_B})e^{-z_AS_A}.
\label{rec2}
\end{eqnarray}
These equations have always the trivial solution $S_A=0, S_B=0$ but depending on the value of the average connectivity in the two networks, $z_A, z_B$, other non trivial solutions might emerge. 
In the following we characterize the phase diagram described by the solution  to the Eqs. $(\ref{rec2})$ keeping in mind that in order to draw the phase diagram of the percolation problem we should consider only the stable solutions of  Eqs. $(\ref{rec2})$ as we have widely discussed in \cite{PerAnt}.  Here we summarize the phase diagram in Figure $\ref{ant_ER_ER}$.
\begin{itemize}
\item
{\em Region I
$z_A<1, z_B<1$.}
In this region there is only the solution $S_A=0, S_B=0$ to the Eqs. $(\ref{rec2})$. 
\item
{\em Region II-A
$z_A>1,z_B<\ln(z_A)/(1-1/z_A)$.}
In this regions there is only one stable solution to the percolation problem $S_A>0 S_B=0$ 
\item
{\em Region II-B
$z_B>1,z_A<\ln(z_B)/(1-1/z_B)$.}
In this regions there is only one stable solution to the percolation problem $S_A=0 S_B>0$ 
\item
{\em Region III
$z_A>\ln(z_B)/(1-1/z_B)$ and $z_B>\ln(z_A)/(1-1/z_A)$.}
In this region we observe two stable solutions of the percolation problem  with  $S_A>0$, $S_B=0$ and $S_A=0, S_B>0$. Therefore in this region we observe a bistability of the percolation configurations.
\end{itemize}
We observe that in this case for each steady state configurations, only one of the two networks can be percolating also in the region in which we observe  a bistability of the solutions.

\section { Percolation on interdependent networks with a fraction $q$ of antagonistic nodes}

In this section we explore the percolation phase diagram when we allow for a combination of antagonistic and interdependent nodes.
As in the previous case we  consider two networks of $N$ nodes. We call the networks, network A and network B respectively and every node $i$ is represented in both networks.

If we indicate by $S_{A} (S_B)$ the probability that a random node in network A (network B) belongs to the percolating cluster in network A (network B), and if we indicate by $S_A^{\prime}(S_B^{\prime})$ the probability that following a link in network A (network B) we reach a node in the percolating cluster of network A (network B), we have
  \begin{eqnarray}
\hspace*{-6mm}S_A&=&q[1-G_0^A(1-S^{\prime}_A)]G_0^B(1-S^{\prime}_B)+\nonumber \\
\hspace*{-6mm}&&+(1-q)[1-G_0^A(1-S^{\prime}_A)][1-G_0^B(1-S_B^{\prime})],\nonumber\\
\hspace*{-6mm}S_B&=&q[1-G_0^B(1-S_B^{\prime})]G_0^A(1-S_A^{\prime})+\nonumber \\
\hspace*{-6mm}&&+(1-q)[1-G_0^B(1-S_B^{\prime})][1-G_0^A(1-S_A^{\prime})].
\label{recS2}
\end{eqnarray}

In the same time, in a random networks with local tree structure  the probabilities   $S_A^{\prime}$ and $S_B^{\prime}$  satisfy the following recursive equations
  \begin{eqnarray}
\hspace*{-6mm}S_A^{\prime}&=&q[1-G_1^A(1-S^{\prime}_A)]G_0^B(1-S^{\prime}_B)+\nonumber \\
\hspace*{-6mm}&&+(1-q)[1-G_1^A(1-S^{\prime}_A)][1-G_0^B(1-S_B^{\prime})],\nonumber\\
\hspace*{-6mm}S_B^{\prime}&=&q[1-G_1^B(1-S_B^{\prime})]G_0^A(1-S_A^{\prime})+\nonumber \\
\hspace*{-6mm}&&+(1-q)[1-G_1^B(1-S_B^{\prime})][1-G_0^A(1-S_A^{\prime})].
\label{rec2}
\end{eqnarray}

\subsection{Two Poisson networks}
We will consider the case of two interacting Poisson networks with average connectivities $z_A=\avg{k}_A$ and $z_B=\avg{k}_B$.
We have seen that for the case of two fully antagonistic Poisson networks the stable percolation configurations correspond to states in which either one of the two networks is percolating.
Therefore with purely antagonistic interactions the system is not able to sustain the coexistence of two percolating clusters present in both  networks.
Here we want to generalize the above case to two interacting networks with only a fraction $q$ of antagonistic interactions.
For two Poisson networks we have $G_0^{A}(x)=G_1^A(x)=e^{z_A(x-1)}$ and $G_0^{B}(x)=G_1^B(x)=e^{z_B(x-1)}$ and therefore $S_A=S_A^{\prime}$ and $S_B=S_B^{\prime}$.
\begin{figure}
\center
\includegraphics[width=0.8\columnwidth]{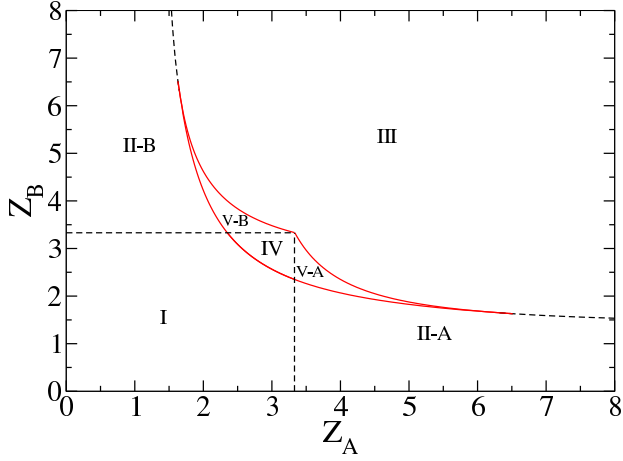}
\caption{Phase diagram two Poisson interdependent networks with a fraction $q=0.3$ of antagonistic interactions.}
\label{phase1}
\end{figure}
\begin{table}
\centering
\begin{tabular}{|l|l|}
\hline
Region I   & $S_A=S_B=0$\\
Region II-A &$S_A>0, S_B=0$\\
Region II-B & $S_A=0,S_B>0$\\
Region III &$S_A>0, S_B>0$ \\
Region IV & $S_A=S_B=0$ and $S_A>0, S_B>0$ \\
Region V-A & $S_A>0,S_B=0$ and $S_A>0, S_B>0$ \\
Region V-B & $S_A=0,S_B>0$ and $S_A>0, S_B>0$ \\
\hline
\end{tabular}
\caption{Stable phases in the different regions of the phase diagram of  the percolation on two antagonistic Poisson networks with a fraction $q=0.3$ of antagonistic nodes (Figure $\ref{phase1}$)}
\label{t0.3}
\end{table}
The Eqs. $(\ref{rec2}), (\ref{recS2})$ can be explicitly written in terms of the average connectivities of the two networks $z_A, z_B$ as

\begin{eqnarray}
S_A&=&f_A(S_A,S_B)=\nonumber\\
&=&(1-e^{-z_A S_A})[(2q-1)e^{-z_B S_B}+1-q]\nonumber \\
S_B&=&f_B(S_A,S_B)=\nonumber\\
&=&(1-e^{-z_B S_B})[(2q-1)e^{-z_A S_A}+1-q]
\label{sol}
\end{eqnarray}
\begin{figure}
\center
\includegraphics[width=1.0\columnwidth]{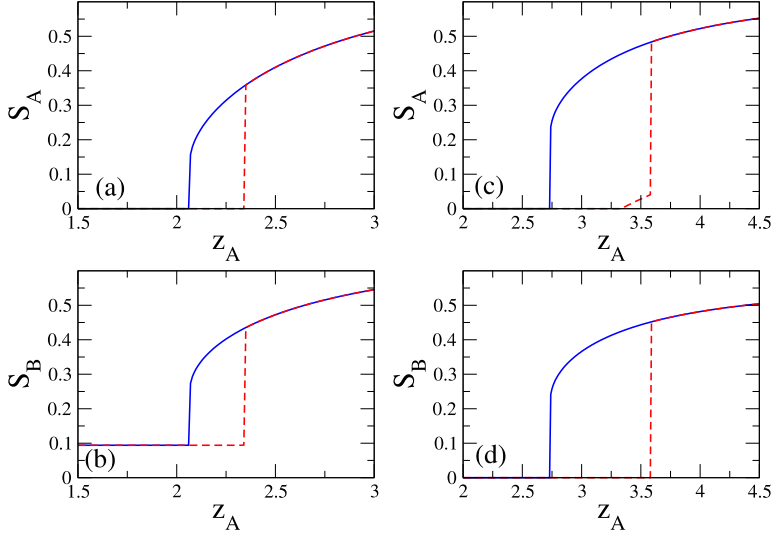}
\caption{(Color online)  Hysteresis loop for $q=0.3$.The hysteresis loop is performed using the method explained in the main text.  The value of the parameter $\epsilon$ used in this figure  is $\epsilon=10^{-3}$. In panel (a) and (b) $z_B=4.0$. In panel (c) and (d)  $z_B=2.8$.}
\label{hysteresis1}
\end{figure}

The solutions to the recursive Eqs.~$(\ref{sol})$  can be classified into three categories:
\begin{itemize}
\item
{\it (i) }The trivial solution in which neither of the network is percolating
$S_A=S_B=0$. 
\item
{\it {(ii)} }The solutions in which just one network is percolating.
In this case we  have either  $S_A>0,S_B=0$ or $S_A=0,S_B>0$. 
From  Eqs.~$(\ref{sol})$ we find that the solution $S_A>0,S_B=0$ emerges  at a critical line of second order phase transition,  characterized by the condition
\begin{equation}
z_A=\frac{1}{q}
\label{c1}
\end{equation}
Similarly the solution $S_B>0,S_A=0$ emerges at a second order phase transition when we have 
\begin{equation}
z_B=\frac{1}{q}.
\label{c2}
\end{equation}
Therefore we observe the phases where  just one network percolates, as long as $q>0$.
This is a major difference with respect to the phase diagram (Figure $\ref{inter_ER_ER}$) of two purelly interdependent networks.
The critical lines Eqs. $(\ref{c1})$ and $(\ref{c2})$ are indicated as dot-dashed lines in the phase diagrams of the percolation transition for different value of the fraction of antagonistic interactions $q$.
\item
{\it {(iii)} }The solutions  for which both networks are percolating. In this case we have $S_A>0,S_B>0$.
This solution can either emerge  (a) when the curves $S_A=f_A(S_A,S_B)$  and $S_B=f_B(S_A,S_B)$ cross at at point $S_A=0$ or $S_B=0$  (b) when the curves $S_A=f_A(S_A,S_B)$  and $S_B=f_B(S_A,S_B)$ cross at a point $S_A\neq 0$  and $S_B\neq 0$ where the two curves are tangent one another. 

For situation (a) the critical line can be determined by imposing, for example, $S_A\rightarrow 0$  in Eqs.~(\ref{rec}), which yields
\begin{eqnarray}
z_B&=&\psi(z_A,q)\nonumber \\
&=&-\frac{\ln{\left(\left[\frac{1}{z_A}-(1-q)\right]/(2q-1)\right)}}{q\left(1-\left[\frac{1}{z_A}-(1-q)\right]/(2q-1)\right)}.
\label{crit2}
\end{eqnarray}
{The function $\psi(z_A,q)$ for $q<0.5$ is a decreasing function of $z_A$ defined for $1/(1-q)<z_A<1/q$, for $q>0.5$ is an increasing function of $z_A$ defined for $1/q<z_A<1/(1-q)$. For $q=0.5$ the function $\psi(z_A,q)$ is not defined but has limit $\psi(z_A,q)\to 1/q=2$.}
\begin{figure}
\center
\includegraphics[width=0.8\columnwidth]{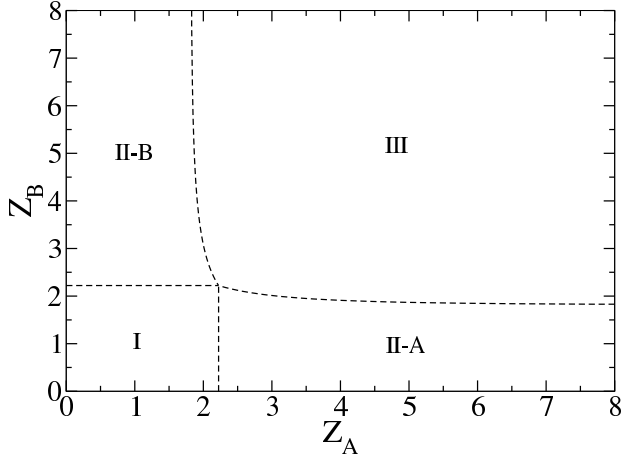}
\caption{Phase diagram two Poisson interdependent networks with a fraction $q=0.45$ of antagonistic interactions.}
\label{phase2}
\end{figure}

\begin{table}
\centering
\begin{tabular}{|l|l|}
\hline
Region I   & $S_A=S_B=0$\\
Region II-A &$S_A>0, S_B=0$\\
Region II-B & $S_A=0,S_B>0$\\
Region III &$S_A>0, S_B>0$ \\
\hline
\end{tabular}
\caption{Stable phases in the different regions of the phase diagram of  the percolation on two antagonistic Poisson networks with a fraction $q=0.45$ of antagonistic nodes (Figure $\ref{phase2}$).}
\label{t0.45}
\end{table}

A  condition similar to Eq. $(\ref{crit2})$ can be found for $z_A, z_B$ by using Eqs.~$(\ref{sol})$ and imposing $S_B\rightarrow 0$.
In particular we obtain the other critical line
\begin{eqnarray}
z_A&=&\psi(z_B,q).
\label{crit1}
\end{eqnarray}

For situation (b) the critical line can be determined imposing that the curves $S_A=f_A(S_A,S_B)$ and $S_B=f_B(S_A,S_B)$, are tangent to each other at the point where they intercept. This condition can be written as
\begin{equation}
\left(\frac{\partial{f_A}}{\partial S_A}-1\right) \left(\frac{\partial{f_B}}{\partial S_B}-1\right)-\frac{\partial{f_A}}{\partial S_B}\frac{\partial{f_B}}{\partial S_A}=0,
\label{tangent}
\end{equation}
where $S_A,S_B$ must satisfy the Eqs.~(\ref{sol}). This is the equation that determines the critical line of first-order phase transition points.

The  condition for having a tricritical point is that Eq.($\ref{crit2}$) or Eq. $(\ref{crit1})$ are satisfied together with  Eq.~$(\ref{tangent})$. If we impose that both Eq.~$(\ref{crit2})$ and Eq.~$(\ref{tangent})$ are satisfied at the same point,   the average connectivities $z_A$ and $z_B$ must satisfy the following conditions
\begin{eqnarray}
 z_B&=&\psi(z_A,q)\nonumber\\
 z_B&=&\phi(z_A,q)=\nonumber \\
 &=&\frac{z_A(2q-1)}{[1-z_A(1-q)][2qz_A(2q-1)+2-3q]}\nonumber \\
\label{tricritical1}
\end{eqnarray}
If we impose that both Eq. $(\ref{crit1})$ and Eq. $(\ref{tangent})$ are satisfied at the same point, the average connectivities $z_A$ and $z_B$ must satisfy the following conditions
\begin{eqnarray}
 z_A&=&\psi(z_B,q)\nonumber\\
z_A&=&\phi(z_B,q)
\label{tricritical2}
\end{eqnarray}
\begin{figure}
\center
\includegraphics[width=0.8\columnwidth]{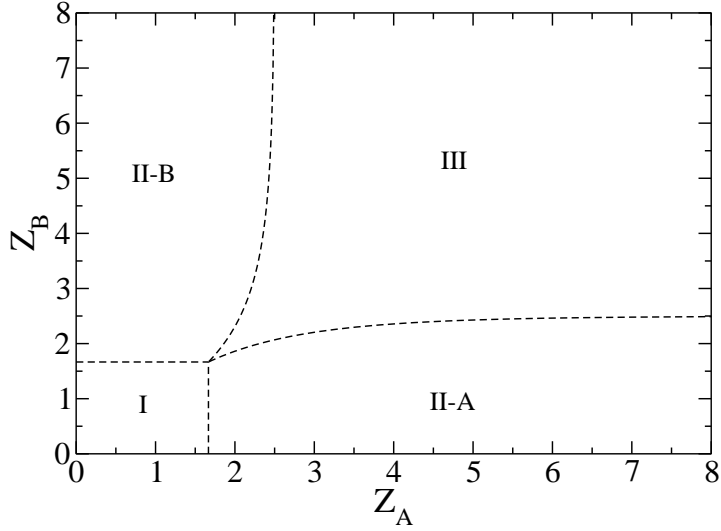}
\caption{Phase diagram two Poisson interdependent networks with a fraction $q=0.6$ of antagonistic interactions.}
\label{phase3}
\end{figure}

In general the systems of Eqs. $(\ref{tricritical1})$ and Eqs. $(\ref{tricritical2})$  have at most two solutions each.
One trivial solution to Eqs.$(\ref{tricritical1})$ and Eqs. $(\ref{tricritical2})$ is $z_A=z_B=\frac{1}{q}$ corresponding to $S_A=S_B=0$.  
In the following we will characterize the solutions to Eqs. $(\ref{tricritical1})$ as a function of the fraction of the antagonist interactions $q$. Similar results can be drawn by studying the system of Eqs. $(\ref{tricritical2})$.
\begin{itemize}
\item
{\it Case $q<0.4$.}
The system of Eqs. $(\ref{tricritical1})$ has two solutions, the trivial solution $z_A=z_B=\frac{1}{q}$ and another non-trivial solution with $z_A<\frac{1}{q}$.  
\item
{\it Case $q=0.4$.}
 The system of Eqs. $(\ref{tricritical1})$ has only one trivial solution with $z_A=z_B=\frac{1}{q}$.Therefore the non-trivial tricritical point disappear.
\item
{\it Case $0.4<q<0.5$.}
The system of Eqs. $(\ref{tricritical1})$ has two solutions, the trivial solution $z_A=z_B=\frac{1}{q}$ and another non-trivial solution with $z_A>\frac{1}{q}$. It turns out that this point is not physical because it is in the region in which the coexistence phase $S_A>0$ and $S_B>0$ cannot be sustained by the system. Therefore in this region we do not have a non-trivial tricritical point. 
\item
{\it Case $0.5<q\le\frac{2}{3}$.} 
The system of Eqs. $(\ref{tricritical1})$ has only the trivial solution $z_A=z_B=\frac{1}{q}$. Therefore the non-trivial tricritical point disappear.
\item
{\it Case $q>\frac{2}{3}$. }
The system of Eqs. $(\ref{tricritical1})$ has two solutions, the trivial solutions $z_A=z_B=\frac{1}{q}$ and another non-trivial solution with $z_A>\frac{1}{q}$.
\end{itemize}

\end{itemize}

\begin{table}
\centering
\begin{tabular}{|l|l|}
\hline
Region I   & $S_A=S_B=0$\\
Region II-A &$S_A>0, S_B=0$\\
Region II-B & $S_A=0,S_B>0$\\
Region III &$S_A>0, S_B>0$ \\
\hline
\end{tabular}
\caption{Stable phases in the different regions of the phase diagram of  the percolation on two antagonistic Poisson networks with a fraction $q=0.6$ of antagonistic nodes (Figure $\ref{phase3}$)}
\label{t0.6}
\end{table}

\begin{figure}
\center
\includegraphics[width=0.80\columnwidth]{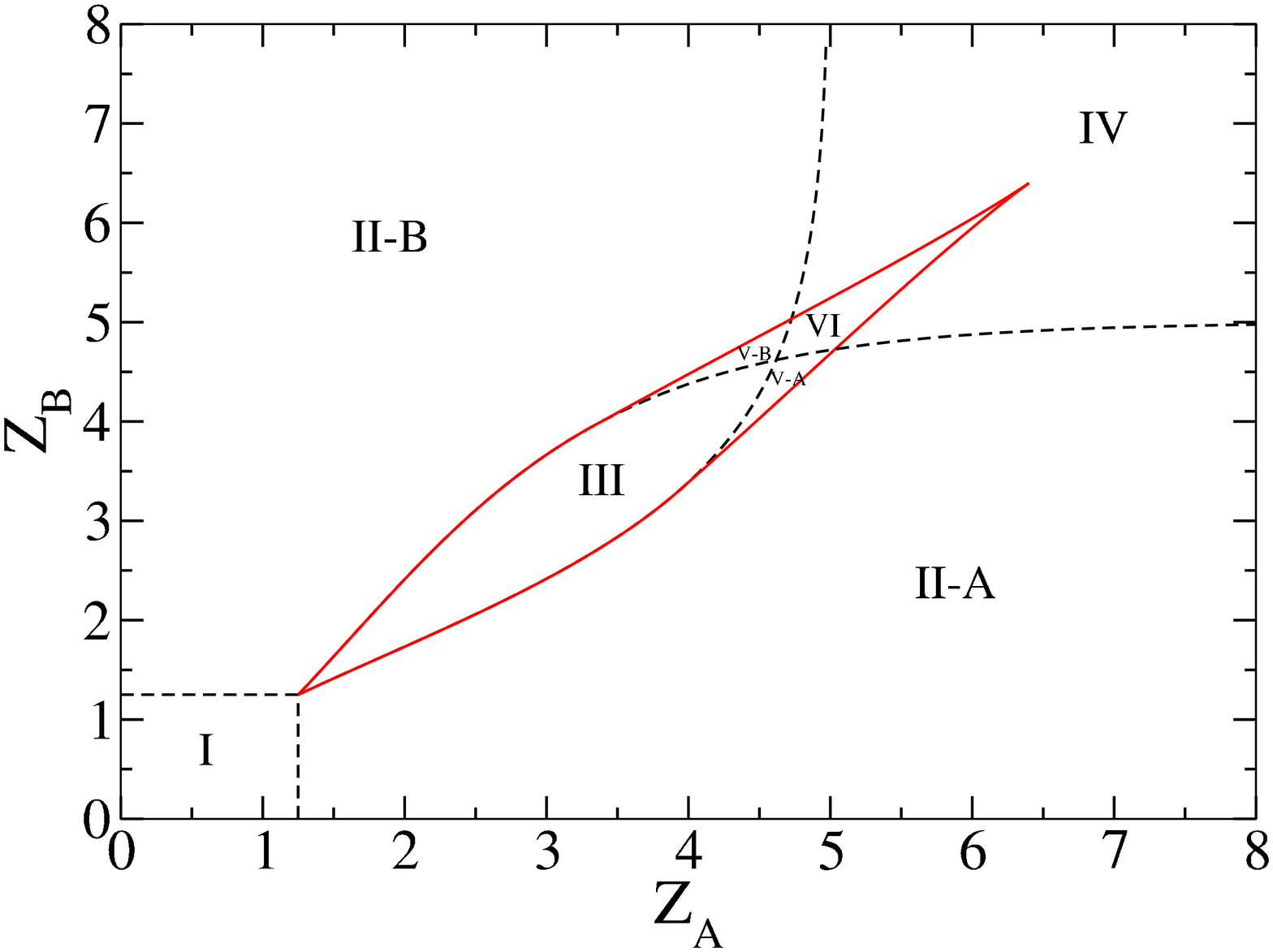}
\caption{Phase diagram two Poisson interdependent networks with a fraction $q=0.8$ of antagonistic interactions.}
\label{phase4}
\end{figure}

\subsection{The phase diagram as a function of $q$ }

\begin{table}
\centering
\begin{tabular}{|l|l|}
\hline
Region I   & $S_A=S_B=0$\\
Region II-A &$S_A>0, S_B=0$\\
Region II-B & $S_A=0,S_B>0$\\
Region III &$S_A>0, S_B=0$ and $S_A=0, S_B>0$\\
Region IV & $S_A>0, S_B>0$ \\
Region V-A & $S_A>0,S_B=0$ and $S_A>0, S_B>0$ \\
Region V-B & $S_A=0,S_B>0$ and $S_A>0, S_B>0$ \\
Region VI & $S_A>S_B>0$ and $S_B>S_A>0$\\
\hline
\end{tabular}
\caption{Stable phases in the different regions of the phase diagram of  the percolation on two antagonistic Poisson networks with a fraction $q=0.8$ of antagonistic nodes (Figure $\ref{phase4}$).}
\label{t0.8}
\end{table} 

\begin{figure}
\center
\includegraphics[width=1.0\columnwidth]{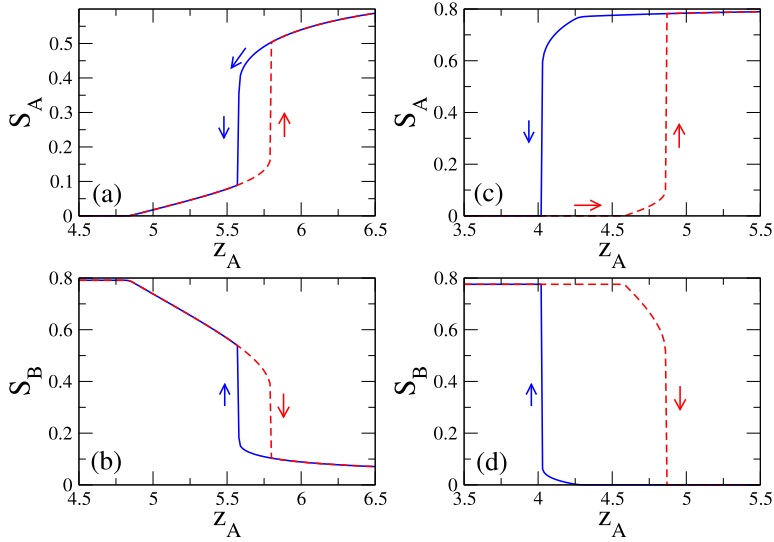}
\caption{(Color online)  Hysteresis loop for $q=0.8$. The hysteresis loop is performed using the method explained in the main text.  The value of the parameter $\epsilon$ used in this figure  is $\epsilon=10^{-3}$. In panel (a) and (b)  $z_B=5.7$. In panel (c) and (d)  $z_B=4.5$.}
\label{hysteresis2}
\end{figure}

As a function of the number of antagonistic interactions $q$ the phase diagram of the percolation problem change significantly. In the phase diagrams reported in Figures $\ref{phase1},\ref{phase2},\ref{phase3},\ref{phase4}$ we plot as red lines the curves along which a first order phase transition can be observed and as black dashed lines the critical lines for a second order phase transition.
\begin{itemize}
\item
{\it Case $q<0.4$.}\\
In Figure $\ref{phase1}$ we show the  phase diagram for $q=0.3$ which is a typical phase diagram in the region $0<q<0.4$. The stable phases in the different regions of the phase space 
are characterized in Table $\ref{t0.3}$. From this table it is evident that in regions IV, V-A and V-B we observe a bistability of the solutions.   {When $q\rightarrow 0$, region II-A, II-B, III, V-A and V-B disappear, reducing the phase diagram to Figure \ref{inter_ER_ER}}.

In order to demonstrate the bistability of the percolation solution in region IV and V-A, V-B  of the phase diagram we solved recursively the Eqs. $(\ref{sol})$ for $z_B=4.0$ (or $z_B=2.8$) and variable values of  $z_A$ (see Figure~\ref{hysteresis1}). We start from values of $z_A=3$, and we  solve recursively the  Eqs.~$(\ref{sol})$. We find the solutions  $S_A=S_A(z_A=3)>0$, $S_B=S_B(z_A=3)=0$. Then we lower slightly $z_A$  and we solve again the Eqs.~$(\ref{sol})$ recursively, starting from the initial condition {$S_A^{ o}=S_A(z_A=3)+\epsilon$}, $S_B^{ o}=S_B(z_A=3)+\epsilon$, and  plot the result.  (The small perturbation $\epsilon>0$ is necessary in order not to end up with the trivial solution $S_A=0,S_B=0$.) Using this procedure we show that if we first lower the value of $z_A$  and then again we raise it,  as shown in Figure~$\ref{hysteresis1}$, the solution present  an hysteresis loop.
This means that in the region IV and V-A, V-B there is a bistability of the solutions.
\item
{\it Case  {$0.4\le q<0.5$}.}\\
In Figure $\ref{phase2}$ we show the phase diagram for $q=0.45$ which is a typical phase diagram in the range $0.4<q<0.5$.
The stable phases in the different regions of the phase space are characterized in Table $\ref{t0.45}$.
For this range of parameters  we do not observe a bistability of the solutions. 
\item{\it Case  {$0.5<q \le \frac{2}{3}$}.}\\
In Figure $\ref{phase3}$ we show the phase diagram for $q=0.6$ which is a typical phase diagram in the range $0.5<q<\frac{2}{3}$. The stable phases in the different regions of phase space are characterized in Table $\ref{t0.6}$.
From this table it is evident that in this case we do not observe bistability of the solutions.
Moreover from the phase diagram Figure $\ref{phase3}$ it is clear that also if the majority of the nodes are antagonistic the interdependent nodes are enough  to sustain a phase in which both networks are percolating at the same time (Region III).

\item
{\it Case $q>\frac{2}{3}$.}\\
In Figure $\ref{phase4}$ we show the phase diagram for $q=0.8$ which is a typical phase diagram in the range $q>\frac{2}{3}$.
In Table $\ref{t0.8}$ we characterize the stable phases in the different regions of the phase diagram. 
Region III, V-A,V-B and VI show a bistability of the solutions. In Figure $\ref{hysteresis2}$ we show evidence that in these regions we can observe an hysteresis loop if we proceed by calculating $S_A,$ and $S_B$ recursively  from Eqs. $(\ref{sol})$ using the same technique used to produce Figure $\ref{hysteresis1}$. For $q>\frac{2}{3}$ the regions in phase space where we observe the coexistence of two percolating phases (Region IV, V-A, V-B and VI) are   reduced and disappear  as $q\to 1$. 

\end{itemize}

\section{Conclusions}
In this paper we  have investigated  how much interdependencies and incompatibilities modify the stability of complex networks and change the phase diagram of the percolation transition.
We found that interdependent networks are  robust against antagonistic interactions, and that we need a fraction $q>q_c=2/3$ of antagonistic interactions for reducing significantly the  region in phase-space in which both networks are percolating.
Nevertheless, we observe that even a small fractions of antagonistic nodes $0<q<0.4$ might induce a bistability of the percolation solutions.
In the future we plan to extend this model to more than two networks,   including a combinatorial complexity \cite{Weigt} of dependency types to cope with the challenges of an increasingly interconnected set of technological, social and economical networks.

\end{document}